\documentclass{mem}
\usepackage{natbib}\usepackage{txfonts}\usepackage{balance}
\usepackage{graphicx}
\usepackage{hyperref}

\def\Sec{\hbox{${}^{\prime\prime}$\llap{.}}}
\def\kms{$\mathrm{km\;s}^{-1}$} 
 
\def\ha{H$\alpha$}

\def\niig{[N~{\small II}]$\,\lambda6583$} 
\def\niipg{[N~{\small II}]$\,\lambda\lambda6548,6583$}

\def\msun{M$_{\odot}$} 
\def\mbh{${\cal M}_{\bullet}$}

\def\mlstar{$(M/L)_\star$}
\def\mlsun{(M/L)$_\odot$}
\def\lbulge{$L_{\it bulge}$}

\def\h3{$h_{3}$}
\def\h4{$h_{4}$}

\idline{75}{282}
\begin{document}
\def\teff{$T\rm_{eff }$}
\def\kms{$\mathrm {km s}^{-1}$}

\title{The Black Hole Mass of \\ Abell~1836-BCG and Abell~3565-BCG}

\author{
  E. \,Dalla Bont\`a\inst{1,2}, 
L. \, Ferrarese\inst{2},
E. M. \, Corsini\inst{1},
J. \, Miralda-Escud\'e\inst{3},
L. \, Coccato\inst{4},
\and
A. \, Pizzella\inst{1}}

  \offprints{E. Dalla Bont\`a}

\institute{
Dipartimento di Astronomia, Universit\`a degli Studi di Padova,
 Vicolo
dell'Osservatorio, 3,
I-35122 Padova, Italy,
\email{elena.dallabonta@unipd.it}
\and Herzberg Institute of Astrophysics, Victoria,
Canada
\and Institut de Ciencies de l'Espai (CSIC-IEEC)/ICREA, Bellaterra, Spain
\and  Max-Planck-Institut fuer extraterrestrische Physik, Garching bei Muenchen, Germany}

\authorrunning{Dalla Bont\`a}

\titlerunning{Supermassive Black Holes in BCGs}

\abstract{
Two brightest cluster galaxies (BCGs), namely
Abell~1836-BCG and Abell~3565-BCG, were observed with the Advanced Camera for
Surveys (ACS) and the Space Telescope Imaging Spectrograph (STIS) on
board the Hubble Space
Telescope. By modeling the available photometric and kinematic data,
it resulted that the mass of Abell~1836-BCG and Abell~3565-BCG are  
$M_\bullet=4.8^{+0.8}_{-0.7}\times 10^9 $ M$_\odot$ and 
$M_\bullet=1.3^{+0.3}_{-0.4}\times 10^9 $ M$_\odot$ at
$1\sigma$ confidence level, respectively.

}
\maketitle{}

\section{Introduction}
Many nearby galaxies have revealed large dark masses in small regions,
so far explained only in terms of supermassive black holes (SBHs, see
for a review Ferrarese \& Ford 2005).  Spectroscopic and photometric
data at high spatial resolution have made it possible to derive
relatively accurate SBH masses, ${\cal M}_{\bullet}$, for a number of
galaxies.  These measurements are reliable if the SBH sphere of
influence is solved.  It was also found that ${\cal M}_{\bullet}$
correlates with other galaxy properties, such as bulge luminosity
(Magorrian et al. 1998, Marconi \& Hunt 2003), light concentration
(Graham et al. 2001), and bulge velocity dispersion $\sigma$
(Ferrarese \& Merritt 2000; Gebhardt et al. 2000).
The inextricable link between the evolution of SBHs and the
hierarchical build-up of galaxies is most directly probed by the
systems that have undergone the most extensive and protracted history
of merging, namely the massive galaxies associated with the largest
SBHs. This is the case of brightest cluster galaxies (BCGs).  Their
large masses and  luminosities, as well as their having a merging
history which is unmatched by galaxies in less crowded environments,
make BCGs the most promising hosts of the most massive SBHs
in the local Universe. Therefore from Laine et al. (2003) it was
selected a sample of
BCGs,
predicted to host the most massive SBHs (based on the
$M_{\bullet}-\sigma$ and $M_{\bullet}- M_{B}$ relations), and for which
the SBH sphere of influence could be well resolved using the 0\Sec1
wide slit of Space Telescope Imaging Spectrograph (STIS) on board the
Hubble Space Telescope (HST).

\begin{figure*}[t!]
\resizebox{\hsize}{!}{\includegraphics[clip=true]{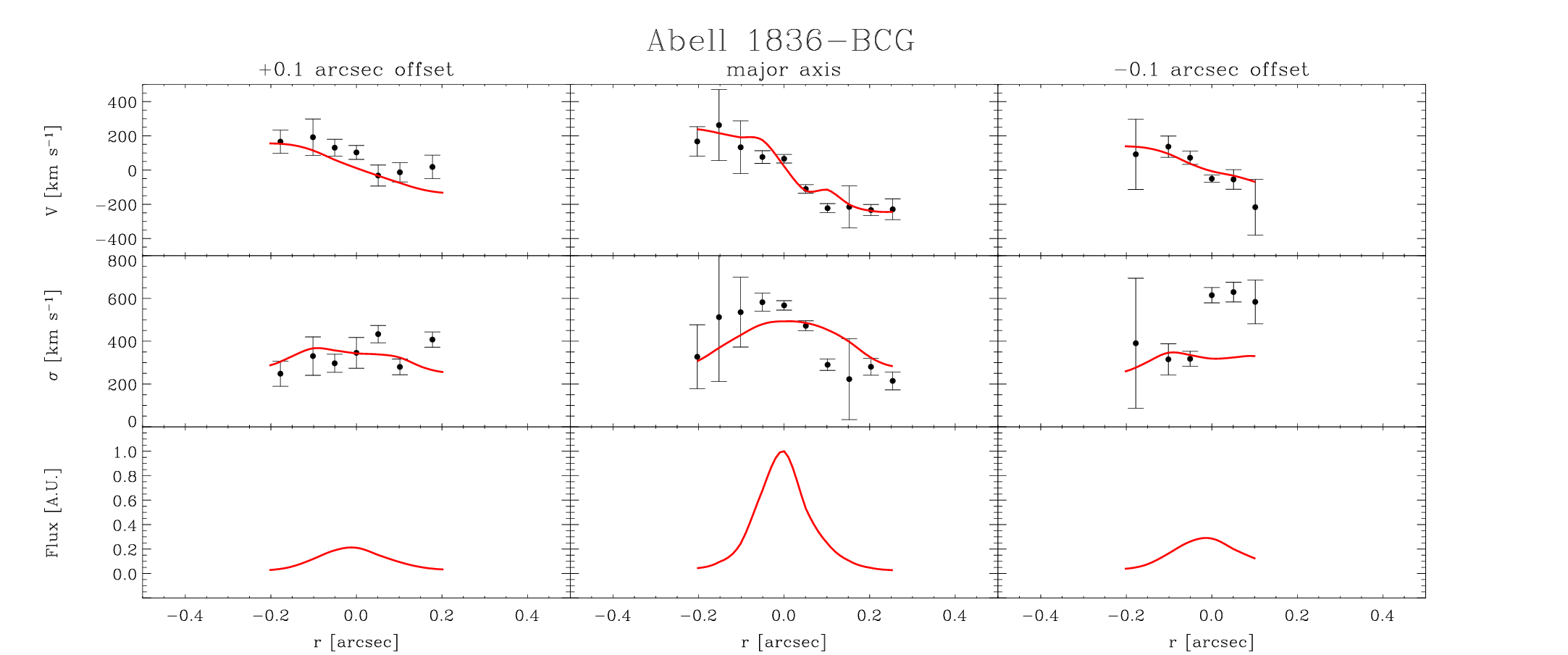}}
\resizebox{\hsize}{!}{\includegraphics[clip=true]{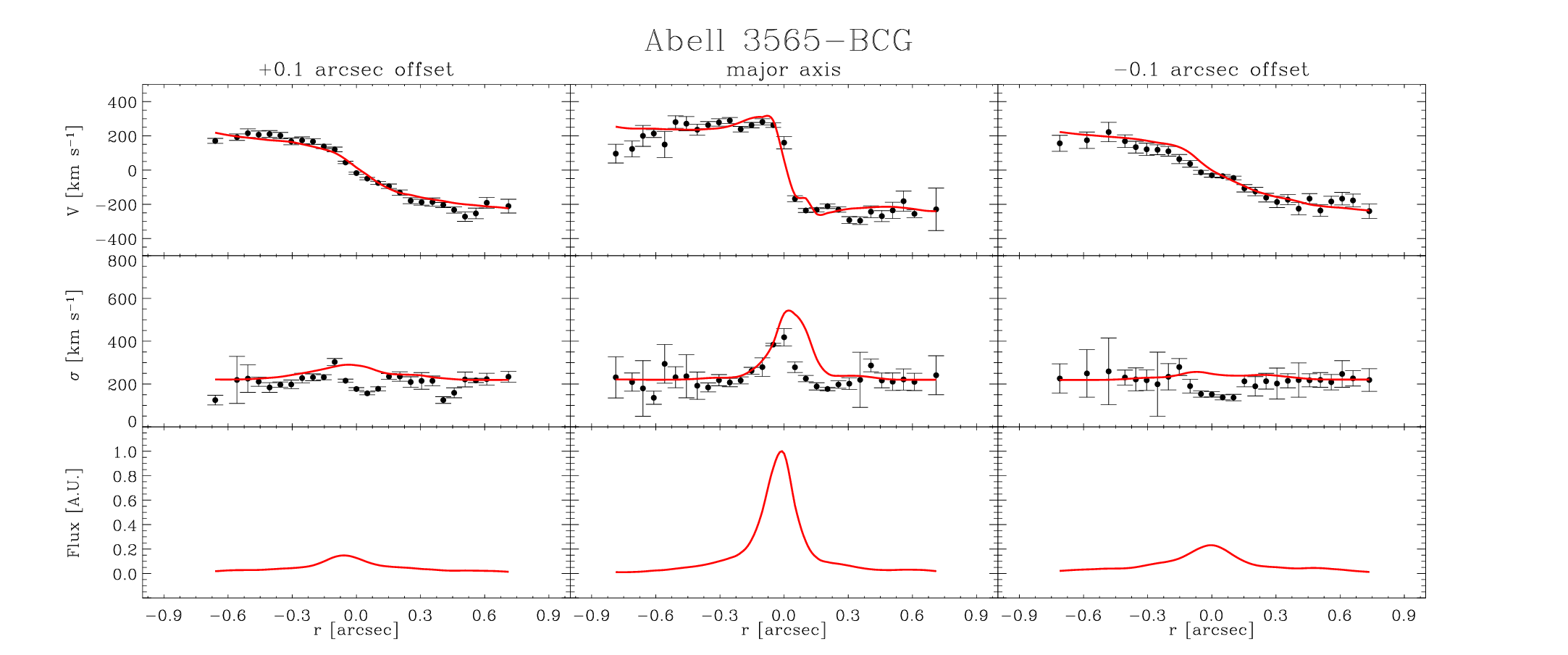}}
\caption{\footnotesize
Observed \niig\ kinematics ({\it filled circles\/}) along
with the best-fitting model ({\it solid line\/}) for the SBH mass of
Abell~1836-BCG and
Abell~3565-BCG.
The observed and modeled velocity curve ({\em top panels}) and velocity dispersion
radial profile ({\em central panels}) are shown for the slit along the
major axis ({\em central panels}), and the two offset positions ({\em left and right panels}) of the gas disk. The corresponding modeled flux profile is shown in the
{\it bottom panels}.
}
\label{obs_mod}
\end{figure*}
\begin{figure}[t!]
\resizebox{\hsize}{!}{\includegraphics[clip=true]{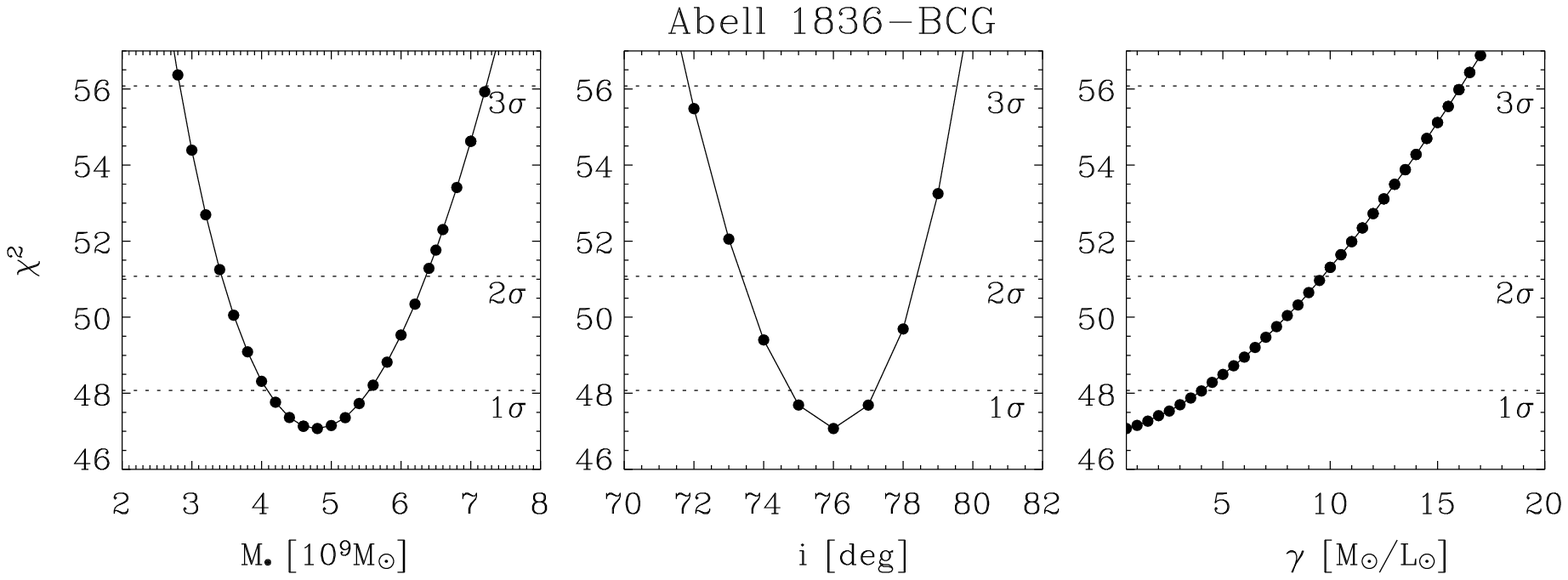}}
\resizebox{\hsize}{!}{\includegraphics[clip=true]{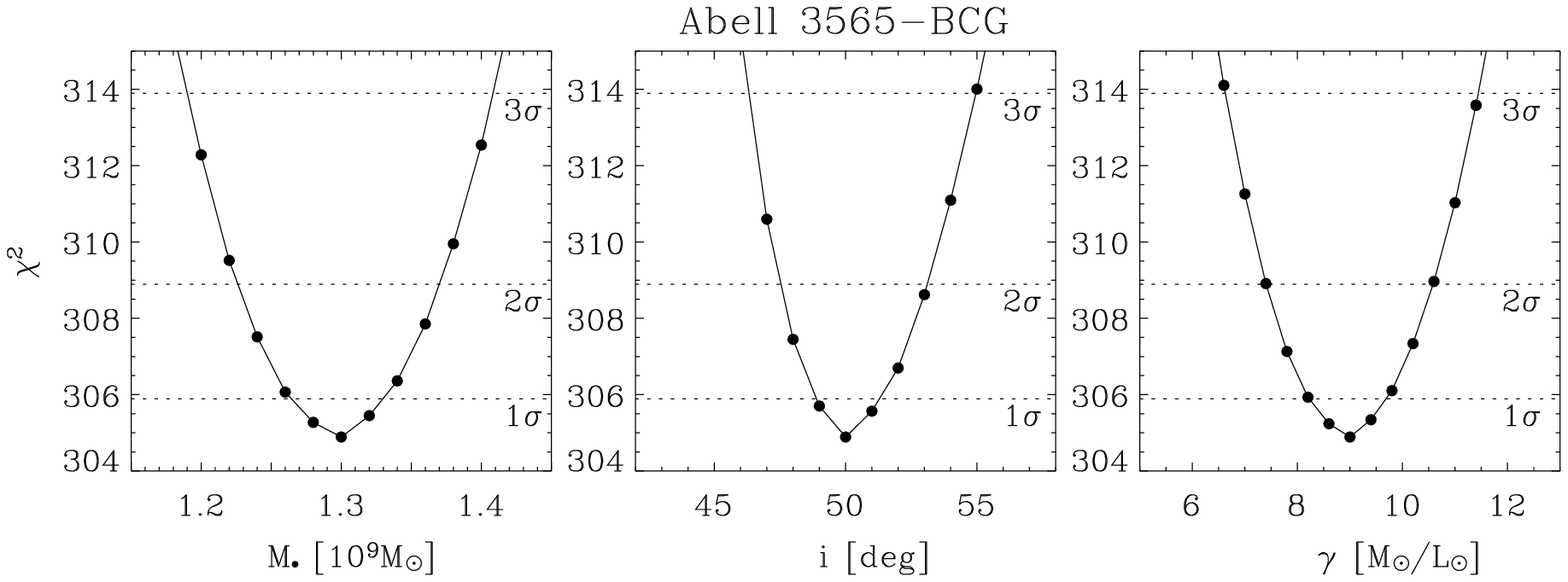}}
\caption{\footnotesize 
$\chi^2$ distribution for Abell~1836-BCG ({\em top panels})  and
Abell~3565-BCG ({\em bottom panels}) as a
    function of SBH mass ({\em left panel}), inclination ({\em central
    panel}), and stellar mass-to-light ratio ({\em right panel}). The dotted horizontal lines indicate the confidence levels on the best fitting values.
}
\label{sigma}
\end{figure}

\section{Imaging and spectroscopy}
Abell~1836-BCG and Abell~3565-BCG were observed with three filters of
the Advanced Camera for Surveys (ACS): F435W (which resembles the
Johnson $B$ filter), F625W (similar to $r$ in the Sloan Digital Sky
Survey photometric system), and the narrow-band ($\approx 130$ \AA)
ramp filter FR656N, covering the redshifted \ha\ and \niipg\ emission
lines in the nucleus of each galaxy. The images were used to determine
the optical depth of the dust, stellar mass distribution near the
nucleus, and intensity map.  Besides, high-resolution spectroscopy of
the H$\alpha$ and [N~{\small II}]$ \,\lambda6583$ emission lines was
obtained for both galaxies, to measure the central ionized-gas
kinematics.  A major axis spectrum was obtained with the slit crossing
the nucleus of the galaxy. Two additional offset spectra were obtained
displacing the slit by one slit width on either side, perpendicularly
to the slit axis. Ionized-gas kinematics was measured by fitting
position, width, and flux of the [N~{\small II}]$ \,\lambda6583$ line.
Figs. 1 and 2 show the kinematics of Abell~1836-BCG and
Abell~3565-BCG, respectively.  They reveal regular rotation curves and
strong central velocity gradients, thus they are good candidates for
dynamical modeling.

\section{Dynamical modeling}
For this study, the procedure described in Coccato et al. 2006, to
which the reader is referred for details, was followed to model the
ionized-gas kinematics.  Briefly, a synthetic velocity field was
generated assuming that the ionized gas is moving in circular orbits
in an infinitesimally thin disk centered at the nuclear location,
under the combined gravitational potential of stars and SBH.  The
model is projected onto the plane of the sky for a given inclination
of the gaseous disk, and then degraded to simulate the actual setup of
the spectroscopic observations.  The latter step includes accounting
for width and location (namely position angle and offset with respect
to the center) of each slit, STIS PSF, and charge bleeding between
adjacent CCD pixels. The free parameters of the model are the mass
\mbh\ of the SBH, mass-to-light ratio \mlstar\ of the stellar
component (and dark matter halo), and inclination $i$ of the gaseous
disk. Both \mlstar\ and $i$ are assumed to be radially invariant.
Although $i$ can be estimated from the images of the disk, the ionized
gas is concentrated in the innermost region of the dust disk and
slight warps can be present even in the case of disks with a regular
elliptical outline (e.g., Ferrarese et al. 1996; Sarzi et al. 2001;
Shapiro et al. 2006). Therefore $i$ is best treated as a free
parameter.  The surface
brightness distribution of the ionized gas is treated as input, by the
narrow-band imaging. The
velocity dispersion of the emission lines was assumed to be described
by the radial function $\sigma(r) = \sigma_0 + \sigma_1
e^{-r/r_\sigma}$, with the choice of the best parameters to
reproduce the observables.  \mbh, \mlstar, and $i$ were determined by
finding the model parameters which produce the best match to the
observed velocity curve, obtained by minimizing $\chi^2 = \sum (v -
v_{\it mod})^2/\delta^2(v)$ where $v\pm\delta(v)$ and $v_{\it mod}$
are the observed and the corresponding model velocity along the
different slit positions, respectively.

\section{Results}

For Abell~1836-BCG a three-dimensional grid of models was explored, with $0 \leq M_\bullet
\leq 3.0\times10^{10}$ \msun~ in 2.0$\times10^8$ \msun\ steps, $
0^\circ \leq i \leq 90^\circ$ in $1^\circ$ steps, and $ 0\leq
(M/L)_\star \leq 40$ \mlsun~ in 0.5 \mlsun\ steps.
The best model fitting the observed rotation curve requires
$M_\bullet=4.8^{+0.8}_{-0.7}\times 10^9$ \msun, $i=76\pm 1^\circ$, and
\mlstar\ $\leq 4.0$ \mlsun, where the errors on $M_\bullet$ and $i$,
and the upper limit on \mlstar, are quoted at the $1\sigma$ confidence
level.  This is the largest SBH mass to have been dynamically measured
to-date.  The model is compared to the observed \niig\ kinematics in
Fig.~\ref{obs_mod}.  Figure~\ref{sigma} shows $1\sigma$,
$2\sigma$, and $3\sigma$ confidence levels individually on \mbh, $i$,
and \mlstar, according to the $\Delta\chi^2$ variations expected for
one parameter, with the other two held fixed at their best-fitting
values.

 The rotation curves of Abell 3565-BCG along the three slit positions
were fitted for a grid of model parameters defined by $0 \leq
M_\bullet \leq 5.0\times10^{9}$ \msun~ in 2.0$\times10^7$ \msun\
steps, $ 0^\circ \leq i \leq 90^\circ$ in $1^\circ$ steps, and $ 0\leq
(M/L)_\star \leq 15$ \mlsun\ in 0.4 \mlsun\ steps.
The best model requires $M_\bullet=1.3^{+0.3}_{-0.4}\times 10^9$
\msun, $i=50\pm 1^\circ$ and \mlstar $=9.0\pm0.8$ \mlsun, where all
errors are given at the $1\sigma$ confidence level.  The model is
compared to the observed kinematics in
Fig.~\ref{obs_mod}. Fig.~\ref{sigma} shows the confidence
levels on \mbh, $i$, and \mlstar\ alone.

\begin{figure}[t!]
\resizebox{\hsize}{!}{\includegraphics[clip=true]{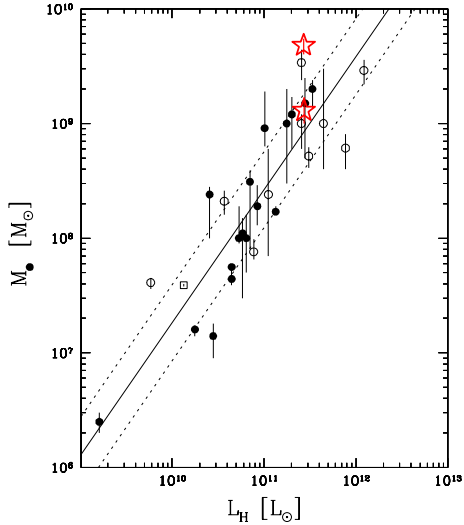}}
\caption{\footnotesize Location of the SBHs masses of our BCGs ({\em
stars}) with respect to the near-infrared \mbh$-$\lbulge\ relation by
Marconi \& Hunt (2003).  NGC 1399 was added (Houghton et al. 2006).
The SBH masses based on resolved dynamical studies of ionized gas
({\it open circles\/}), water masers ({\it open squares\/}), and stars
({\it filled circles\/}) are plotted.  Dotted lines represent the
$1\sigma$ scatter of the relation.}
\label{scaling}
\end{figure}

\section{Conclusions}

Several authors have recently focused their attention on the position
of BCGs and massive galaxies in the upper end of the SBH scaling
relations, using both theoretical and, even if indirect, observational
considerations.  Bernardi et al. (2007) found that BCGs define a
shallower $\sigma-{\cal L}$ relation than the bulk of early-type
galaxies. They interpreted it as due to a non-linear correlation
between log\mbh\ and log$\sigma$. This would bring the \mbh$-\sigma$
to underestimate \mbh\ for high $\sigma$ values. Furthermore,
according to von der Linden et al. (2007) BCGs follow a steeper
Faber-Jackson relation than non-BCGs. This implies that BCGs can
follow at most one of the power-law relations, either the
\mbh$-\sigma$ or the \mbh$-${\cal L}.  Both Bernardi et al. (2007) and
von der Linden et al. (2007) used SDSS data.  On the contrary,
Batcheldor et al. (2007) argue that SBHs masses predicted from NIR
luminosities are consistent with masses predicted from $\sigma$. They
attributed the discrepancies to the presence of extended blue
envelopes around the BCGs.

In order to compare the SBH mass determinations of Abell 1836-BCG and
Abell 3565-BCG with the prediction of the near-infrared
\mbh$-$\lbulge\ relation by Marconi \& Hunt (2003) the 2MASS $H$-band
luminosities of the two BCGs were retrieved from the NASA/IPAC
Infrared Science Archive.  Figure~\ref{scaling} shows the location of
the SBH masses in the \mbh$-$\lbulge\ plane. Only the SBH mass of
Abell 1836-BCG is not consistent with the relation, since it is larger
than the one expected (i.e., $8.9\times10^8$ \msun).  The same is true
only for M 87, among the seven galaxies with \mbh$> 10^9$ \msun\
dynamically measured so far.  The $\sigma$ of Abell~1836-BCG is not
available in literature, thus it is not possible to discuss its
behavior in the \mbh$-\sigma$ relation. On the other hand, all the
galaxies with \mbh$> 10^9$ \msun\ lie on the relation.  There is no
strong evidence that the upper end of the scaling relations is
regulated by a different law.

The only way to hush-up the debate is adding new \mbh.  Before the
advent of new classes of telescopes from earth and space, adaptive
optics is a viable solution (e.g., Houghton et al. 2006).  It will be
possible to understand the behavior of both \mbh$-\sigma$ and
\mbh$-{\cal L}$ relations and, in case, if BCGs and normal galaxies
have a bimodal trend.

\bibliographystyle{aa}

\end{document}